\documentclass[aps,twocolumn,showpacs,amsmath,amssymb,prx]{revtex4-1}

\usepackage{graphicx}

\usepackage{comment}
\usepackage[normalem]{ulem}
\usepackage{bm}

\renewcommand{\thefigure}{\arabic{figure}}

\makeatletter
\def\blfootnote{\xdef\@thefnmark{}\@footnotetext}
\makeatother

\usepackage{color}
\usepackage{ulem}

\begin{document}
\renewcommand{\figurename}{\textbf{Fig.}}
\renewcommand{\thefigure}{\textbf{\arabic{figure}}}
%
\title{
Unconventional thermal metallic state of  charge-neutral fermions in an insulator
}

\author{ Y. Sato$^1$, Z. Xiang$^2$, Y. Kasahara$^1$,  T. Taniguchi$^1$, S. Kasahara$^1$,  L. Chen$^2$, T. Asaba$^2$, C. Tinsman$^2$, H. Murayama$^1$, O. Tanaka$^3$, Y. Mizukami$^3$,  T. Shibauchi$^3$, F. Iga$^4$, J. Singleton$^5$}
\author{Lu Li$^2$}
\email{luli@umich.edu}
 \author{Y. Matsuda$^1$}
\email{matsuda@scphys.kyoto-u.ac.jp}

\affiliation{
$^1$Department of Physics, Kyoto University, Kyoto 606-8502, Japan\\
$^2$Department of Physics, University of Michigan, Ann Arbor, MI  48109, USA\\
$^3$ Department of Advanced Materials Science, University of Tokyo, Chiba 277-8561, Japan\\
$^4$Institute of Quantum Beam Science, Graduate School of Science and Engineering, Ibaraki University, Mito 310-8512, Japan\\
$^5$Los Alamos National Laboratory, Los Alamos, NM 87545
}
\date{\today}

\maketitle

{\bf Quantum oscillations (QOs) in transport and thermodynamic parameters at high magnetic fields are an unambiguous signature of the Fermi surface, the defining characteristic of a metal. Therefore, recent observations of QOs in insulating SmB$_6$ \cite{Li2014,Tan,Hartstein,Xiang2017} and YbB$_{12}$, in particular the QOs of the resistivity $\rho_{xx}$ in YbB$_{12}$ \cite{Xiang2018},  have been a big surprise, pointing to the formation of a novel state of quantum matter. Despite the large charge gap inferred from the insulating behaviour of $\rho_{xx}$, these compounds seemingly host a Fermi surface at high magnetic fields. However, the nature of the ground state in zero field has been little explored.  Here we report the use of low-temperature heat-transport measurements to discover gapless, itinerant, charge-neutral excitations in the ground state of YbB$_{12}$.  At zero field, despite  $\rho_{xx}$ being far larger than that of conventional metals, a sizable linear temperature dependent term in the thermal conductivity is clearly resolved in the zero-temperature limit ($\kappa_{xx}/T(T\rightarrow 0)=\kappa_{xx}^0/T\neq 0$), analogous to normal metallic behaviour. Such a residual $\kappa_{xx}^0/T$ term at zero field, which is absent in SmB$_6$ \cite{Hartstein,Xu2016,Boulanger},  leads to a spectacular violation of the Wiedemann-Franz law: the Lorenz ratio $L=\kappa_{xx}\rho_{xx}/T$ is $10^{4}$--$10^{5}$ times larger than that expected in conventional metals. These data indicate that YbB$_{12}$ is a charge insulator but a thermal metal, suggesting the presence of itinerant neutral fermions.  Remarkably, more insulating crystals with larger activation energies exhibit  a larger amplitude of the resistive QOs as well as a larger $\kappa_{xx}^0/T$, in stark contrast to conventional metals.   Moreover, we find that these fermions  couple to magnetic field, despite their charge neutrality. Our findings expose novel gapless and highly itinerant, charge-neutral quasiparticles in this unconventional quantum state. }


In intermetallic $4f$ and $5f$ compounds, strong hybridization between itinerant and predominately localized electrons often opens an insulating gap \cite{Tsunetsugu,Riseborough}.   Among such {\it Kondo insulators}, SmB$_6$ and YbB$_{12}$ have recently aroused great interest due to several remarkable properties.  Theoretical work suggests that  both are topological  insulators \cite{Dzero2010,Weng2014}, which host three dimensional (3D) insulating bulk and metallic 2D surfaces.  The surface states in SmB$_{6}$ are protected by time reversal and inversion symmetries, while those in YbB$_{12}$ are protected by crystal symmetry.   In both compounds, the metallic surface states have been established using a number of  experimental techniques, including angle-resolved photoemission spectroscopy (ARPES) \cite{Xu2014,Hagiwara}.   In particular, in SmB$_6$,  spin-resolved ARPES has shown the spin-momentum locked surface, which is a characteristic feature of a topological insulator.

Recently another salient aspect of both compounds has come as a great surprise.    In an external magnetic field, SmB$_6$ exhibits quantum oscillations (QOs) in magnetization (the de Haas-van Alphen [dHvA] effect), which is associated with Landau quantization~\cite{Li2014,Tan,Hartstein,Xiang2017,Xiang2018}.   However, it is still unclear as to whether the dHvA signal in SmB$_6$ results from the metallic surface or  insulating bulk states.    Moreover, it has been pointed out theoretically that QOs in magnetization can indeed occur in a certain type of band insulator via magnetic breakdown in strong magnetic fields~\cite{Knolle2015,Knolle2017}.  Even more exotic possibilities have been suggested, such as neutral fermions that form a Fermi surface~\cite{Baskaran,Erten,Chowdhury,Sodemann} or the consequences of non-Hermitian properties in strongly correlated systems~\cite{Yoshida2018,Shen2018}.  Following these scenarios, unusual quasiparticles, such as  composite excitons and neutral Majorana fermions, have been proposed.  If these charge-neutral degrees of freedom form structures similar to the Fermi surface of metals,  they may well produce dHvA oscillations.

\begin{figure*}[t]
	\begin{center}
		\includegraphics[width=0.8\linewidth]{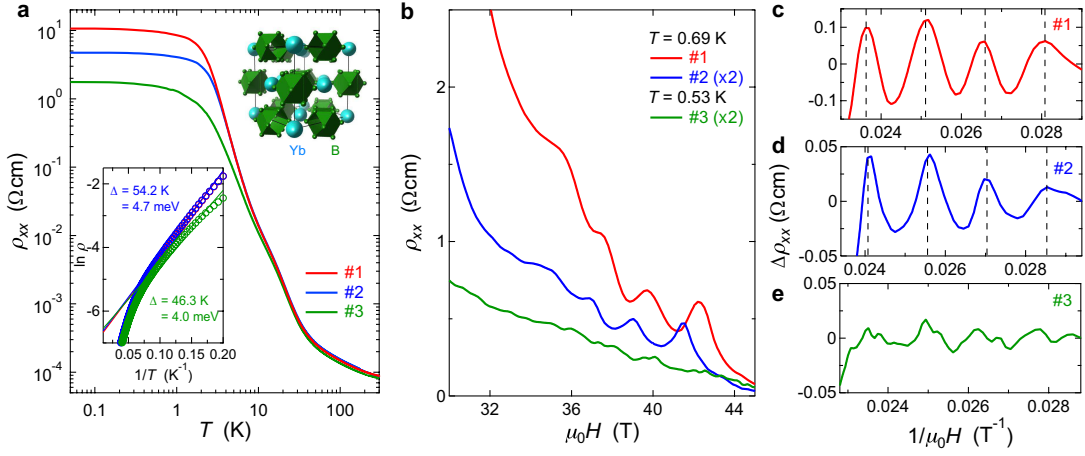}
		\caption{{\bf Resistivity of YbB$_{12}$.} {\bf a,} Temperature dependence of resistivity in three crystals (\#1 and \#2 from the same batch and \#3 from another batch) of YbB$_{12}$. The inset is an Arrhenius plot above 5\,K. Also shown is the crystal structure;     YbB$_{12}$ crystallizes in the face-centred cubic UB$_{12}$-type structure, in which strongly covalently bonded units of B$_{12}$ cubooctahedra form cubic lattice and interstitial Yb atoms are accommodated in the octahedral pores among these units.  {\bf b,} Field dependence of the resistivity at high magnetic fields applied close to the $c$ axis at 0.69\,K for \#1 and \#2 and 0.53\,K  for \#3. {\bf c-e,} The oscillatory part of magnetoresistance plotted against the inverse field in crystals \#1 ({\bf c}), \#2 ({\bf d}) and \#3 ({\bf e}). }
		\label{fig:figure1}
	\end{center}
\end{figure*}

A natural consequence of the Fermi surface is the linear temperature dependence of the heat capacity $C$ and thermal conductivity $\kappa_{xx}$ at low temperatures.  While $C$ includes both localized and itinerant excitations, $\kappa_{xx}$ is determined exclusively by itinerant excitations.   In SmB$_6$, although a finite linear heat-capacity coefficient $\gamma$ is observed~\cite{Tan},  there is no term in the zero-field thermal conductivity that is linear in $T$ ({\it i.e.} $\kappa_{xx}^0/T=0$)~\cite{Hartstein,Xu2016,Boulanger}.  This suggests that itinerant gapless neutral fermions are absent in zero field. There has been intense debate as to whether the itinerant neutral fermions are excited by magnetic field;  whilst field-induced enhancement of the thermal conductivity has been attributed to excitations of neutral fermions \cite{Hartstein}, an alternative interpretation involving conventional phonon mechanisms has been pointed out~\cite{Boulanger}.  In addition, neutron inelastic scattering experiments  reveal distinct excitation modes within the hybridization gap \cite{Fuhrman}, but there is no evidence of charge-neutral excitations.   Whether or not there are nontrivial itinerant quasiparticles, which may be responsible for the observed dHvA oscillations,  therefore remains a controversial issue in the case of SmB$_6$.

The very recent discovery of resistivity QOs (the Shubnikov-de Haas [SdH] effect) in another Kondo insulator YbB$_{12}$ (SdH oscillations are not observed in SmB$_6$),   has revealed a novel aspect of the QOs in insulators \cite{Xiang2018}.  In YbB$_{12}$, $4f$ and $5d$ band hybridization leads to a narrow insulating gap \cite{Mignot,Okawa,Terashima}, with the mean valence of the Yb ions being close to $+3$ ($4f^{13}$ state)~\cite{Yamaguchi}.  Compared to the situation in the mixed-valence compound SmB$_6$, this suggests a simpler electronic state in YbB$_{12}$, with the $f$-electrons mostly localized. 

The 3D nature of the SdH signal in YbB$_{12}$ demonstrates that the QOs in the resistivity arise from the electrically insulating bulk \cite{Xiang2018}.  In addition, the QOs in YbB$_{12}$ behave in other ways that are different from those in SmB$_6$.   In SmB$_6$, the effective masses  $m^*$ of the quasiparticles determined from dHvA oscillations are much smaller than the free electron mass $m_e$, indicating that 
 correlation effects are of little importance.  Moreover, the temperature dependences of the QO amplitudes in some SmB$_6$ crystals deviate strikingly from the predictions of the standard Lifshitz-Kosevich (LK) formula applicable to Fermi liquids at very low temperatures \cite{Tan,Hartstein}. By contrast, in YbB$_{12}$, the $m^*$ values estimated from SdH oscillations are much larger than $m_e$,  implying 	strong correlation effects.  Moreover, the oscillations accurately obey the LK formula, showing no deviation from Fermi-liquid theory \cite{Xiang2018}.

Figure\,1a depicts the $T$-dependence of the resistivity $\rho_{xx}$  for three different crystals (crystals \#1 and \#2 are taken from the same batch and \#3 is from a different batch) grown by the floating zone technique.  In all crystals, $\rho_{xx}$ increases by four to five orders of magnitude from room temperature to 0.1\,K.  Below $\sim2$\,K, $\rho_{xx}$ becomes weakly temperature-dependent, resembling the $T<3.5$\,K resistive ``plateau" well known in SmB$_6$~\cite{Cooley}; this is attributed to the topological metallic surface state. The residual resistivities $\rho^0_{xx}$ are approximately 11, 4.5 and 1.8\,$\Omega$cm for crystals \#1, \#2 and \#3, respectively.  The inset of Fig.\,1a shows an Arrhenius plot of the resistivity above 5\,K, where the surface conduction is negligible.  Obviously $\rho_{xx}$ of all crystals show an activation-type temperature dependence with two-gap behaviour.  The resistivities of \#1 and \#2 overlap above 5\,K, while that of \#3 is lower than \#1 and \#2 below $\sim20$\,K.   Fitting with a thermal activation model of the resistivity ($\rho_{xx}(T) \propto \exp(\Delta/2k_BT)$) we obtain gap widths of 4.7\,meV for \#1 and \#2  and 4.0\,meV for \#3 over the temperature range 6\,K $< T <$ 12.5\,K. 
Despite their similar activation energies, $\rho^0_{xx}$ of crystal \#1 is 2.2 times larger than that  of \#2 at the lowest temperatures. This is consistent with the presence of the 2D metallic surface.   In fact, assuming the same surface conductance of both crystals,   the $\rho_{xx}^0$ obtained from the surface area of \#2 single crystal is roughly 2 times smaller  than that of \#1.

Figure\,1b depicts the field dependence of the resistivity  $\rho_{xx}(H)$  at  0.69\,K for \#1 and \#2 and at 0.53\,K for \#3 with the magnetic field applied close to the $c$ axis.  Upon applying field, the negative slope of the $\rho_{xx}(T)$ curve is preserved up to 45\,T with no signature of metallic behaviour, indicating that the energy gap still remains \cite{Xiang2018}.   Figures\,1c, d and e display the  oscillatory part of the resistivity   $\Delta \rho_{xx}$, which is obtained by subtracting a polynomial background from $\rho_{xx}(H)$, plotted as a function of $1/\mu_0H$.   For crystals \#1 and \#2,  four periods with an approximately constant spacing  provide strong evidence that these are SdH oscillations. The small difference in the peak positions between \#1 and \#2 is due to a slight difference in field direction.  A direct quantitative comparison of the carrier scattering rate between crystals \#1 and \#2 is difficult due to the different surface contributions, but the fact that the oscillations start above approximately 33\,T in both crystals suggests similar scattering rates, which is consistent with the samples possessing the same activation energy.  For crystal \#3, on the other hand, no discernible oscillations  are observed (Figs.\,1b and e), suggesting a larger scattering rate.   In insulators, a large activation energy usually indicates a low impurity concentration and high crystallographic quality.  From this point of view, the quality of crystals \#1 and \#2 is similar, and better than \#3 with its lower activation energy.  Thus, remarkably,  the more insulating crystals exhibit larger SdH oscillations, the opposite of the behaviour of conventional metals.

\begin{figure}[t]
	\begin{center}
		\includegraphics[width=1.0\linewidth]{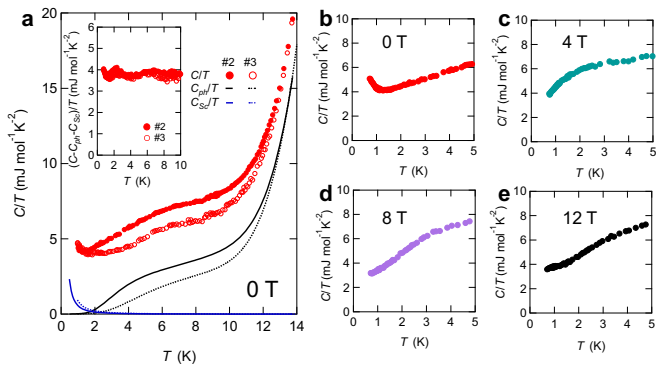}
		\caption{{\bf Heat capacity of YbB$_{12}$.} {\bf a,} Temperature dependence of heat capacity divided by temperature $C/T$ in zero field for \#2 and \#3 crystals.  The black solid and dotted lines represent the phonon heat capacity obtained by Debye 
		and Einstein 
		contributions (see Methods). The best fit is obtained by $\beta=0.026$\,mJ/molK$^4$ and two Einstein modes with $\Theta_E=16$ and 170\,K for \#2 crystal and $\beta=0.017$\,mJ/molK$^4$ and  $\Theta_E=24$ and 160\,K for \#3 crystal.
		The blue solid and dotted lines represent the Schottky contributions obtained by the three level model (see also Methods), 
with $\Delta_1=7.1$ and $\Delta_2=8.3$\,meV, and $\Delta_1=6.8$ and $\Delta_2=7.6$\,meV for \#2 and \#3 crystal, respectively.
		  The inset shows the quasiparticle contribution $C_{qp}/T$ which is obtained by subtracting the Schottky and phonon contributions from the total heat capacity.  {\bf b-e,} $C/T$ of \#2 crystal  at low temperatures in zero field ({\bf b}) and in magnetic fields applied along [1,0,0] direction ({\bf c,d,e}).}
		\label{fig:figure2}
	\end{center}
\end{figure}

Figure\,2a depicts the temperature dependence of the heat capacity divided by temperature ($C/T$) of crystals \#2 and \#3 in zero field.    As shown in Fig.\,2b, $C/T$ shows a slight upturn below $\approx$1\,K, which is attributed to a Schottky contribution $C_{\rm Sc}/T$.  Despite this, it is obvious that an extrapolation of $C/T$ from above 1\,K to $T = 0$ has a finite intercept, indicating the presence of a linear temperature term, {\it  i.e.,} the gapless quasiparticle excitations possess $C_{\rm qp}=\gamma T$.  Thus the heat capacity can be written as a sum of phonon, quasiparticle and Schottky contributions, $C=C_{\rm ph}+C_{\rm qp}+C_{\rm Sc}$.  The low-temperature enhancement of $C/T$ is well fitted by a three-level Schottky model, as shown by blue solid and dotted lines in Fig.\,2a.  As reported in isostructural compounds LuB$_{12}$ and YB$_{12}$, a hump anomaly around 6\,K and steep increase above 10\,K in $C/T$ may be attributed to low-energy optical phonon modes of Yb atoms in the cavities of the B$_{24}$ cuboctahedrons \cite{Czopnik}.   The solid and dotted black lines indicate  $C_{\rm ph}/T$ obtained from an acoustic phonon contribution ($\propto T^3$) and two optical phonon contributions.  The optical phonon contributions are slightly sample dependent.  Owing to the high Debye temperature, the acoustic phonon contribution to the total heat capacity is very small.      As shown in the inset of Fig.\,2a,  $C_{\rm qp}/T$ (obtained by subtracting $C_{\rm ph}$ and $C_{\rm Sc}$ from the total $C$) is in good agreement for crystals \#2 and \#3.  Thus,  we obtain $\gamma\approx 3.8$\,mJ/mol\,K$^2$ in zero field, which is comparable to values in conventional metals. Figures 2c, d and e show $C/T$ in magnetic field.  At $\mu_0H$=4\,T, $C/T$ decreases with decreasing $T$ with a downward curvature below 2\,K.  At $\mu_0H$=8 and 12 \,T, $C/T$ decreases nearly linearly with $T$ with steeper slope than that of the zero-field data.  This low temperature behaviour may be attributed to  the coupling between the magnetic field and the optical phonons; however, a quantitative estimation is difficult.  Nevertheless, it is obvious that a simple extrapolation of $C/T$ to $T = 0$  indicates that $\gamma$ is slightly reduced by magnetic field.

\begin{figure}[t]
	\begin{center}
		\includegraphics[width=1.0\linewidth]{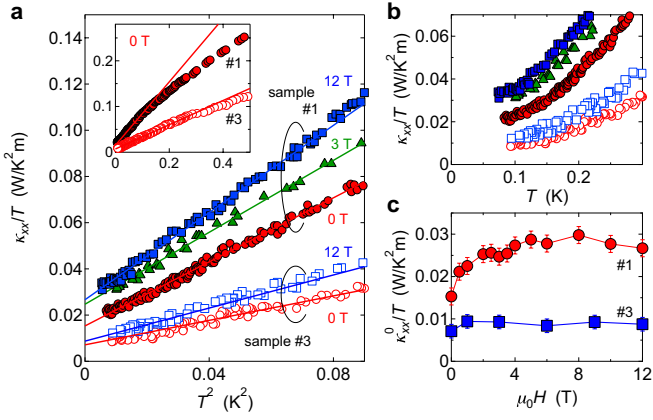}
		\caption{{\bf: Thermal conductivity of YbB$_{12}$.} {\bf a,} Thermal conductivity divided by temperature $\kappa_{xx}/T$ plotted as a function of $T^2$ in zero field and at $\mu_0H=3$ and 12\,T for \#1 and \#3 crystals at low temperatures.    The solid lines represent  $\kappa_{xx}/T=\kappa_{xx}^0/T+AT^2$ obtained by the fitting.   The inset shows the same data up to  $T^2=0.5$\,K$^2$ in zero field.  The deviation form the fitting occurs at $\sim 0.4$\,K for \#1 and $\sim 0.5$\,K for \#3 crystal.    The  {\bf b,} The same data plotted as a function of $T$.   {\bf c,}  Field dependence of residual linear thermal conductivity term $\kappa_{xx}^0/T$ obtained by the extrapolation to zero temperature at each field. }
		\label{fig:figure3}
	\end{center}
\end{figure}

We now turn to the thermal conductivity that shows the itinerant aspect of the neutral excitations.   The filled and open red circles in the inset of Fig.\,3a  depict the $T$-dependence of $\kappa_{xx}/T$ below 0.3\,K in zero field  for crystals \#1 and \#3, respectively.    As the thermal conductivity does not contain the localized Schottky contribution, $\kappa_{xx}$ can be described as a sum of the itinerant quasiparticle and phonon contributions, $\kappa_{xx}=\kappa^{\rm qp}_{xx}+\kappa^{\rm ph}_{xx}$.    In order to use $\kappa_{xx}$ as a probe of itinerant quasiparticles, $\kappa_{xx}^{\rm ph}$ must be extracted reliably.     Figure\,3a depicts $\kappa_{xx}/T$ plotted as a function of $T^2$, revealing that $\kappa_{xx}/T=\kappa_{xx}^0/T+AT^2$ at low temperatures. The $AT^2$-term is attributable to phonons for the following reasons.    In non-magnetic insulators, acoustic phonons are the only carriers of heat at low temperature and the phonon conductivity is given by $\kappa_{xx}^{\rm ph}=\frac{1}{3}\beta T^3 v_{\rm ph}\ell_{\rm ph}$, where $v_{\rm ph}$ and $\ell_{\rm ph}$ are the sound velocity and mean free path of acoustic phonons, respectively.   We compare $\ell_{\rm ph}$ and the effective diameter of the sample $d_{\rm eff}$=$2\sqrt{wt/\pi}$ ($w$ and $t$ are the width and thickness of the crystal, respectively); $d_{\rm eff} = 0.58$~mm and 0.37\,mm for crystals \#1 and \#3, respectively.    Using $\beta=$0.026 and 0.017\,mJ/mol K$^4$ for crystals \#1 and \#3, respectively (obtained from the measurements shown in Fig.\,2a) and  $v_{\rm ph}=9.6\times10^3$\,m/s for LuB$_{12}$ \cite{Grechnev}, we find that $\ell_{\rm ph} \approx d_{\rm eff}$  at $\sim 0.5$\,K for crystal \#1 and $\sim 0.6$\,K for \#3.   These temperatures are close to the temperatures below which $\kappa_{xx}/T$ shows $T^2$-dependence, as shown in the inset of Fig.\,3a, supporting the above estimation.  These results suggest that at low enough temperatures ($T^2 \lesssim 0.1$\,K$^2$) $\ell_{\rm ph}$ will be limited by the crystal size, {\rm i.e.,} the samples are in the boundary-scattering regime where  $\kappa_{xx}^{\rm ph}/T\propto T^2$.  The fact that the systems are in this regime is also supported by the the $A$-values of crystals \#1 and \#3.   In the boundary scattering regime, $A$ is proportional to $\beta d_{\rm eff}$.    The ratio of $A$ values of crystals \#1 and \#3 determined by the $T$-dependence of $\kappa_{xx}/T$  is $\approx 2.6$.  This value is  close to the ratio ($\approx 2.5$) of $\beta d_{\rm eff}$  of the two crystals, indicating the proportionality of $A$ and $\beta d_{\rm eff}$.

As revealed by both plots of  Figs.\,3a and 3b,  $\kappa_{xx}/T$ extrapolated to zero temperature yields definite non-zero intercepts in both crystals, $\kappa^0_{xx}/T \neq 0$.   
This indicates a finite residual linear term in  $\kappa^{\rm qp}_{xx}$, {\it i.e.,} the presence of itinerant gapless excitations.  It should be stressed that the observed finite $\kappa^0_{xx}/T$  does {\it not} originate from charged quasiparticles, in contrast to the situation in conventional metals. Evidence for this is given by the spectacular violation of the Wiedemann-Franz (WF) law, which connects the electronic thermal conductivity $\kappa_{xx}^e$ to the electrical resistivity.   In moderately pure metals at low temperatures,  $L=\kappa_{xx}^e\rho_{xx}/T \leq L_0$ is generally satisfied,  where $L_0=\frac{\pi^2}{3}\left(\frac{k_B}{e}\right)^2=2.44\times10^{-8}$\,W$\Omega$K$^{-2}$ is the Lorenz number~\cite{Singleton}.  The values of $\kappa_{xx}^0\rho^0_{xx}/T$ for crystals \#1 and  \#3 are found to be $\sim6\times 10^4L_0$ and $\sim5\times 10^3L_0$, respectively.  As the surface metallic regime is expected to follow the WF law, our results imply that the neutral fermions in the insulating bulk of the samples are responsible for the finite $\kappa_{xx}^0/T$.  In other words, as the bulk resistivity diverges as $T\rightarrow 0$, the Lorentz number for the heat carrying quasiparticles also diverges.  Thus the thermal conductivity and heat capacity data very strongly suggest the presence of highly mobile and gapless neutral fermion excitations in zero field, which are not observed in SmB$_6$.

We note that $\kappa_{xx}^0/T$ for crystal \#1 is nearly twice as large as that for \#3, while $\gamma$ for \#2, whose quality is very close to \#1, coincides with that for crystal \#3.   The quasiparticle thermal conductivity is related to the heat capacity by
\begin{equation}
\frac{\kappa_{xx}^{\rm qp}}{T}=\frac{1}{3}\gamma v_{\rm F} \ell_{\rm qp},
\end{equation}
where $v_{\rm F}$ is the Fermi velocity and $\ell_{\rm qp}$ is the mean free path of the neutral fermions.  Therefore $\ell_{\rm qp}$ of crystals \#1 and \#2 is twice as large  as that of \#3.  Interestingly, this indicates that more insulating crystals with larger activation energies have higher mobility neutral quasiparticles, supporting the assertion made above when discussing the QOs.

As shown in Fig.\,3a, $\kappa_{xx}/T$ is greatly enhanced by applying magnetic field.   More importantly, as depicted in  Fig.\,3c,  $\kappa_{xx}^0/T$, which is obtained by extrapolating $\kappa_{xx}/T$ to zero temperature at each field, is enhanced by field.  It should be remembered that $\kappa_{xx}^0/T$ contains no phonon contribution.   Therefore, the field-induced enhancement of $\kappa_{xx}^0/T$ implies that the neutral fermions couple to magnetic fields.  Another prominent feature is that $\kappa_{xx}^0/T$ of crystal \#1 is much more enhanced by magnetic field than that of \#3, indicating  that better quality crystals with lower impurity scattering rates exhibit larger magneto-thermal conductivity.  As larger $\kappa_{xx}^0/T$  values arise from longer mean free paths, this result suggests (as might be expected) that the more mobile neutral fermions are more strongly influenced by magnetic field.


A fascinating question is whether the charge neutral fermions are responsible for the QOs.    To examine this, we estimate $\ell_{\rm qp}$ from Eq.(1) by assuming that  $v_{\rm F}$ is given by the Fermi velocity obtained from the SdH oscillations.  By assuming a simple spherical Fermi surface, we obtain $v_{\rm F}=\hbar k_{\rm F}/m^*\approx 1.3\times 10^4$\,m/s  from the SdH oscillations,   where $k_{\rm F}\approx 1.7$\,nm$^{-1}$ is the Fermi wave number and $m^*\approx 15 m_{\rm e}$ is the effective mass \cite{Xiang2018}.   We estimate  $\ell_{\rm qp}\approx$54 and 25\,nm, which is nearly 70 and 30 times longer than the lattice constant, for  crystals \#1 and \#3, respectively.  Although the mean-free path is long,  the heavy effective mass leads to rather small mobilities: the mobility $\mu = \frac{e\ell_{\rm qp}}{m^*v_{\rm F}}$ is about 480 cm$^2$/Vs (0.048\,T$^{-1}$) for crystal \#1 and 230 cm$^2$/Vs  (0.023\,T$^{-1}$)  for \#3. This simple model explains why 30 - 40\,T magnetic fields are needed to resolve the SdH oscillations and why the oscillations in crystal \#3 are much smaller than those in crystal \#1. Therefore, this rather crude estimate suggests that the enhanced thermal conductivity in zero field and the resistive QOs at high fields are intimately connected; {\it i.e.}, the long mean-free paths imply that the neutral fermions  are responsible for the QOs.

\begin{figure}[t]
	\begin{center}
		\includegraphics[width=0.85\linewidth]{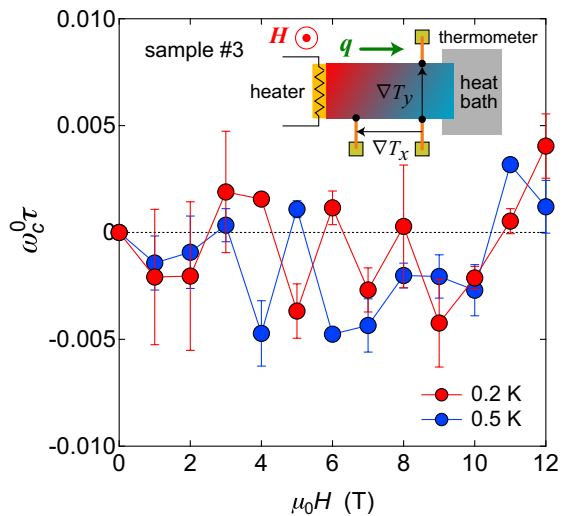}
		\caption{{\bf Thermal Hall angle of YbB$_{12}$.} Field dependence of the thermal Hall angle $\omega_c^0 \tau \equiv \kappa_{xy}/\kappa_{xx}^0$ at 0.2 and 0.5\,K.  Because of $\kappa_{xx}>\kappa_{xx}^0$,  $\omega_c\tau$ is smaller than $\omega_c^0\tau$. The inset illustrates the experimental setup.}
		\label{fig:figure4}
	\end{center}
\end{figure}

Very recently, a thermal Hall effect of neutral fermions that experience the Lorentz force, akin to the conduction electrons in metals, has been proposed \cite{Chowdhury,Katsura}.  In an attempt to observe such an effect, we measured thermal Hall conductivity $\kappa_{xy}$.  According to Ref.\,\cite{Chowdhury,Katsura},  the tangent of the thermal Hall angle,
 \begin{equation}
 \tan \theta_H= \kappa_{xy}/\kappa_{xx}=\omega_{\rm c}\tau,
 \end{equation}
provides similar information on the electrical Hall angle in conventional metals~\cite{Singleton}.  Here, $\omega_{\rm c} =eb/m^*$ corresponds to the cyclotron frequency of the neutral fermion, where $b$ is the effective magnetic field experienced by neutral fermions, and $\tau$ is the scattering time.  Figure\,4 depicts the field dependence of $\omega_c^0 \tau \equiv \kappa_{xy}/\kappa_{xx}^0$ at 0.2 and 0.5\,K.  Because of $\kappa_{xx}>\kappa_{xx}^0$,  $\omega_{\rm c}\tau$ is smaller than $\omega_{\rm c}^0\tau$.  No discernible thermal Hall effect is observed; $\omega_{\rm c}^0\tau$ and hence $\omega_{\rm c}\tau$ is less than 0.005 within our resolution.  In conventional metals, $\omega_{\rm c}\tau=eB\tau/m^*$ becomes order of unity at the magnetic field where the quantum oscillations appear.  

As the SdH oscillations are observed around 40\,T \cite{Xiang2018}, the thermal Hall angle at 10\,T could be expected to be of order 0.2, which is much larger than the observed thermal Hall angle.  However, it is premature to conclude that the neutral fermions are not responsible for the SdH oscillations, because the small thermal Hall angle may be explained by a non-linear $B$-dependence of $b$ or the presence of electron- and hole-like pockets of neutral fermions. In the latter scenario, compensation effects may reduce the thermal Hall signal considerably.


The presence of a Fermi surface of neutral fermions and the  coupling to external magnetic field with negligible thermal Hall angle calls for further studies.
The existence of the itinerant neutral fermions adds another piece to the puzzle of anomalous insulating states with metallic quantum oscillations.

\section*{ACKNOWLEDGMENTS}

We thank  K. Behnia, D. Chowdhury, P. Coleman, J. Knolle, E.-G. Moon,  R. Peters,  S. Sebastian, T. Senthil, and L. Taillefer for fruitful discussions. This work was supported by Grants-in-Aid for Scientific Research (KAKENHI) (Nos. 25220710, 15H02106, 15H03688, 16K13837, 18H01177, 18H01180, 18H05227) and on Innovative Areas ``Topological Material Science" (No. 15H05852) from Japan Society for the Promotion of Science (JSPS).   This work at Michigan is mainly supported by the Office of Naval Research through the Young Investigator Prize under Award No. N00014-15-1-2382 (electrical transport characterization), by the National Science Foundation under Award No. DMR-1707620 (magnetization measurement), by the National Science Foundation Major Research Instrumentation award under No. DMR-1428226 (the equipment of the electrical transport characterizations). The development of the torque magnetometry technique in intense magnetic fields was supported by the Department of Energy under Award No. DE-SC0008110. A portion of this work was performed at the National High Magnetic Field Laboratory, which is supported by National Science Foundation Cooperative Agreement No. DMR-1644779, the Department of Energy (DOE) and the State of Florida. JS thanks the DOE for support from the BES program ``Science in 100~T''. The experiment in NHMFL is funded in part by a QuantEmX grant from ICAM and the Gordon and Betty Moore Foundation through Grant GBMF5305 to Dr. Ziji Xiang, Tomoya Asaba, Lu Chen, Colin Tinsman, and Dr. Lu Li. We are grateful for the assistance of Tim Murphy, Hongwoo Baek, Glover Jones, and Ju-Hyun Park of NHMFL.

\section*{Author contributions}
F.I. grew the high-quality single crystalline samples. Y.S., Y.K., S.K., and H.M. performed the thermal transport measurements. T.T., S.K., O.T., and Y.Mizukami performed the heat capacity measurements. Z.X., L.C, T.A., C.T., J.S., and L.L. performed the high-field resistivity measurements. Y.S., Z.X.,Y.K.,T.T., S.K., H.M., Y.Mizukami, T.S., L.L., and Y.Matsuda analyzed the data.  T.S., J.S., L.L., and Y.Matsuda prepared the manuscript.

\newpage
\noindent {\bf Methods}\\
\noindent {\bf Crystal growth and sample preparation.}
YbB$_{12}$ single crystals were grown by the traveling-solvent floating-zone method \cite{Iga}. Three crystals (\#1, \#2, and \#3) were cut from the as-grown ingot and polished into a rectangular shape. \#1 and \#2 crystals were taken from the same growth batch and \#3 crystal was from a different growth batch. The dimensions of samples are 4.0$\times$0.52$\times$0.51\,mm$^3$ (\#1), 1.0$\times$1.0$\times$0.2\,mm$^3$ (\#2), and 1.8$\times$0.81$\times$0.13\,mm$^3$ (\#3). Crystals \#1, \#2, and \#3 correspond to sample N3, N1, and N4 in ref.\,\cite{Xiang2018}, respectively.

\bigskip



\noindent {\bf High-field resistivity measurements.}
The magnetoresistivity of YbB$_{12}$ samples \#1, \#2 and \#3 were measured in a capacitor-driven 65\,T pulsed magnet in NHMFL, Los Alamos. Low environment temperature was achieved by using a $^3$He cryostat. For different probes we used, the base temperature varied, which was 0.69\,K for \#1 and \#2 crystals and at 0.53\,K for \#3 crystal. Magnetic field was applied parallel to one of the cubic axes. An offset of a few degrees was expected due to the misalignment in sample mounting. Considering the different sizes and geometries of the samples, we measured MR with the current parallel to the magnetic field ($I \parallel B$) in sample \#1, and $I \perp B$ in samples \#2 and \#3. For all three samples, we used standard four-contact configuration and the MR data was taken via a high-frequency ac technique \cite{Kohama} with a specialized digital lock-in program. The driving signal ($f = 192$-256\,kHz) was generated by an ac voltage source and applied to the sample across a transformer. The current through the sample was monitored using a 10 Ohm shunt resistor, and was determined to be 38-66\,$\mu$A in our experiment.

\bigskip

\noindent {\bf Heat capacity measurements.}
Heat capacity is measured down to $T \sim 0.6$\,K by a long-relaxation calorimetry using a bare chip resistance thermometer (Cernox 1030BR, Lakehore)~\cite{Taylor}, which enables very small addenda contribution to the total heat capacity.
The heat capacity of the addenda including the grease is measured before the sample is mounted.
The temperature dependence of the addenda is fitted by a polynomial function and subtracted from the total heat capacity. The obtained heat capacity is analyzed as a sum of phonon, quasiparticle and Schottky contributions, $C = C_{ph} + C_{qp} + C_{Sc}$, where the $C_{ph}$ includes Debye, $C_D=\beta T^3$, and Einstein, $C_E=\frac{A}{T} \left(\frac{\Theta_E}{T}\right)^2\frac{\exp{(\Theta/T)}}{[\exp (\Theta_E/T)-1]^2}$, contributions. The Schottky contribution is obtained by using the three level model, $ C_{Sc} = \frac{B}{T^2}\left\{\frac{\sum^{n-1}_{i=0} \Delta_i^2e^{-\Delta_i/T}}{\sum^{n-1}_{i=0}e^{-\Delta_i/T}} - \left(\frac{\sum^{n-1}_{i=0} \Delta_ie^{-\Delta_i/T}}{\sum^{n-1}_{i=0}e^{-\Delta_i/T}}\right)^2 \right\} $ where $n (=3)$ denotes the number of the energy levels, and $\Delta_n$ is the excitation energy.

\bigskip

\noindent {\bf Thermal and thermal Hall conductivity measurements. }
The longitudinal and transverse thermal conductivities, $\kappa_{xx}$ and $\kappa_{xy}$, are measured by the steady-state method, applying the thermal current $\bm{q}$ 
with 
$\bm{q}\parallel\bm{x}\parallel$ [100] and $\bm{H}\parallel\bm{z}\parallel$ [001]. The thermal gradients $-\nabla_xT \parallel\bm{x}$ and $-\nabla_yT \parallel\bm{y}$ were detected by RuO$_2$ thermometers, and $\kappa_{xx}=w_{xx}/(w_{xx}^2+w_{xy}^2)$ and $\kappa_{xy}=w_{xy}/(w_{xx}^2+w_{xy}^2)$ were obtained from the thermal resistivity $w_{xx}=\nabla_xT/q$ and thermal Hall resistivity $w_{xy}=\nabla_yT/q$. The effect of misalignment of the Hall contacts was eliminated by reversing the magnetic field at each temperature.

\end{document}